\documentclass{JHEP}
\usepackage{epsfig}

\title{Supersymmetric  Multiple Basin Attractors}

\author{Renata Kallosh\thanks{On leave of
absence from Stanford University until 1 September 2000}\\
    Theory Division, CERN CH 1211 Geneve 23, Switzerland\\
    E-mail: \email{kallosh@physics.stanford.edu}}
\author{Andrei Linde\thanks{On leave of
absence from Stanford University until 1 September 2000}\\
    Theory Division, CERN CH 1211 Geneve 23, Switzerland\\
    E-mail: \email{linde@physics.stanford.edu}}
\author{ Marina Shmakova \\
    Department of Physics, Stanford University, Stanford, CA 94305, USA\\
 and University of Tennessee, Knoxville, TN 37996, USA\\
    E-mail: \email{shmakova@slac.stanford.edu}}
\received{\today}       
 \preprint{CERN-TH/99-305\\ \hepth{9910021}}

\abstract{We explain that supersymmetric attractors in general have several
critical points due to the algebraic nature of the stabilization equations.
We show that the critical values of the cosmological constant of the
$adS_5$ vacua are given by the topological (moduli-independent) formulae
analogous to the entropy
 of the $d=5$ supersymmetric black holes.  In one-moduli case  critical points with
positive definite metric and gauge couplings exist under condition that the
central charge changes the sign from one critical point to the other. We
have found several families of ${\bf Z}_2$-symmetric critical points where
the central charge has equal absolute values
 but  opposite signs in two attractor points. We present examples
  of interpolating solutions and discuss their generic  features.}


\begin{document}

1. The concept of supersymmetric attractors with respect to black hole
entropy was introduced in \cite{FKS}. The so-called stabilization equations
\cite{S,FK} for the behaviour of the moduli near the charged extremal black
holes horizon have
 been studied extensively during the last few years. It has been
established that
 the supersymmetric fixed points of the theory correspond to the
minimum of the
 central charge\footnote{The world central charge here is used for
the central charge of the supersymmetry algebra as explained in \cite{FK}.}
in the physical
   part of the moduli space, when the metric is positive-definite
\cite{GFK}.

The issue of uniqueness of the basin of attraction of the supersymmetric
systems related to Calabi-Yau black holes of ungauged supergravity
\cite{GST} and cosmological constant in $d=5$ gauged supergravity
\cite{GST}  was raised in \cite{CKRRSW}. It was  also shown there that the
critical points of these two theories are defined by the same equations.
 It was emphasized
in \cite{moore} that one may look for more than one basin of attraction in
the context of the black hole fixed points. The situation however was not
resolved in general: on the one hand one may expect that the black hole
with given electric and magnetic charges in the theory with
compactification on particular CY space have a unique entropy;  on the
other hand the stabilization equations \cite{FK,Marina,CKRRSW}
\begin{equation}
C_{IJK} \bar t^J \bar t^K=q_I \ , \label{stab}\end{equation} being
algebraic equations,
 may
lead to multiple solutions\footnote{We use the notations of \cite{CKRRSW},
where ${\cal V} \equiv C_{IJK} t^i t^J t^K=1$ and $n$ fields $t^I, I=1,
\dots , n$ are functions of $n-1$ independent moduli $\phi^i, i=1, \dots ,
n-1.$}. The reason is that  one starts with the system of $n$ quadratic
equations (\ref{stab}) for $n$ variables $\bar t^I $. They can be reduced
to some higher order algebraic equation for each of the fixed moduli. For
example, in case $I=1,2$  one has to solve quadratic equations that have 2
solutions, etc.

Indeed,  solutions describing black holes with multiple basins
 of attraction have been studied in \cite{Marina1}. It was found for the 2 moduli
 black holes in $d=4$ that only in  one  basin
of attraction  the scalar and vector fields have positive metric. This
suggested that the argument based on the requirement of uniqueness of
entropy may be correct, and only one basin of attraction is  physically
acceptable. Recently a solution with two basins of attraction of analogous
stabilization equations
 was found in \cite{BC}
in application to anti-de Sitter space $adS_5$, but the issue of positivity
of vector field metric was not analyzed there. As a result, the possibility
to have physically acceptable black hole or $adS_5$ configurations with
multiple basins of attraction remained open.

 Recent developments with AdS/CFT
correspondence
 \cite{malda} and BPS
domain walls \cite{kelly}-\cite{hawking} stimulated us to clarify the
situation with the multiple basins of attraction  for supersymmetric
attractors. The purpose of this paper is to establish the conditions under
which supersymmetric attractors may have more than one physically
acceptable
 basin of attraction. We will also study the configurations interpolating
between two basins.

The reason why these two completely different phenomena, $d=5$ CY black
holes and the cosmological constant of $adS_5$ space,  may be treated
simultaneously {\it at the supersymmetric critical points} was explained in
\cite{CKRRSW}. It is important to stress here however that out of critical
points the systems are different: the main difference is that the term with
$(\partial_i Z)^2$ enters with the opposite sign in the black hole
potential and in the gauge theory potential, as shown below.

To find the black hole entropy we are looking for the supersymmetric
critical points of the black hole potential \cite{GFK},  which for $d=5$
ungauged $N=2$ supergravity is given by
$
V= Z^2+ {3\over 2} g^{ij} \partial_i Z \partial_j Z.
$
 Here the central charge
$
Z = t^I q_I
$
depends on {\it real}  moduli $t^I$ and on electric charges of the black
hole $q_I$. The supersymmetric critical points are at
$
  Z_{,i}=0,
$
where
$
\partial _i V =  4 Z \, \partial _i Z   -  \sqrt 6 \;  T_{ijk}
\partial ^j Z
\partial ^k Z=0,
$
where $  T_{ijk}$ is a function of moduli and $C_{IJK}$. At the
supersymmetric critical point of the central charge, which is also a
supersymmetric critical point of the black hole potential, the value of the
potential defines the BPS mass and the black hole entropy:
\begin{equation}
 M^2_{BPS} = |Z|^2, \qquad  ( M^2_{BPS})_{cr} =
|Z|_{cr}^2(C_{IJK}, q_I)=| V_{cr}|
  \quad  {\rm  at} \; Z_{,i}=0.
\label{critV} \end{equation}

 In the
supersymmetric critical points the second derivative of the black hole
potential is proportional to the metric on the moduli space:
$
V _{,i; j}={8\over 3} g_{ij} V_{\rm cr}$ at $
\partial_i V= \partial_i Z =0. $ For
the positive moduli space metric the potential $V$ has a local minimum
whenever the stabilization equation has a solution with non-vanishing
central charge. When the potential is non-zero at the critical point, it
defines the black hole entropy. The entropy $S= {\pi^2\over 12}\tilde S $
is a function of the critical value of the BPS mass
\begin{equation}
 \tilde S=   ( M^2_{BPS})_{cr}^{3/2} = |Z|_{cr}^{3}(C_{IJK},
q_I)=|
 V_{cr}|^{3/2}.
\end{equation}

Now consider the supersymmetric critical points of the potential of the
$U(1)$ gauged $N=2$ $d=5$ supergravity. The size of the $adS_5$ throat is
defined by
 the extrema of the gauged supergravity potential \cite{GST,CKRRSW}.
The relevant potential
 is equal to  $-  6 P$, where
$
P=  Z^2- {3\over 4}  g^{ij} \partial_i Z \partial_j Z.
$
In the context of gauged supergravity the central charge $Z$ is
 a moduli-dependent combination of gravitino and gaugino charges
$V^I$, which is defined by
$
Z = t^I V_I,
$
where $V_I$ is the charge defining the gravitino--gravitino--vector and
gaugino--gaugino--vector interactions.  The critical points of the $P$ are
given by
$
\partial _i P =  Z \, \partial _i Z   +  \sqrt {3/ 2} \;  T_{ijk}
\partial ^j Z
\partial ^k Z=0,
$
and as before the supersymmetric critical points of the central charge are
also the supersymmetric critical points of the potential. The value of the
potential $P$ at the critical point is given by the square of the BPS mass
as a function of $C_{IJK}$ and $V_I$:
\begin{equation}
 M^2_{BPS} = |Z|^2, \qquad  ( M^2_{BPS})_{cr} =
|Z|_{cr}^2(C_{IJK}, V_I)=
 | P_{cr}| \quad  {\rm  at} \; Z_{,i}=0.
\end{equation}
The $adS_5$ vacua are solutions of this theory with unbroken supersymmetry.
The cosmological constant of the relevant $adS_5$ space coincides with  the
critical value of the  BPS mass extremized in the moduli space. At the
supersymmetric critical points, where $Z_{,i}=0$, one finds
\begin{equation}
 \Lambda_{adS_5}=- 6 | t^I  V_I |^2= -6  |Z_{\rm cr}  (V_I,
C_{IJK})|^2 = - 6 M^2_{BPS} \qquad  {\rm at} \qquad Z_{,i}=0.
\end{equation}
 In the
supersymmetric critical points the second derivative of the potential is
proportional to the metric on the moduli space:
$
-(P) _{,i; j}=- \partial_i  \partial_i P = -{2\over 3} g_{ij} (Z^2)_{\rm
cr}$ at $ \partial_i P= \partial_i Z =0. $ For the positive moduli space
metric the potential $-P$ has a maximum whenever the stabilization equation
has a solution. When the potential is non-zero at the critical point, it
defines the cosmological constant. Thus, as explained in Sec. 4.2 of
\cite{CKRRSW} the critical points of the BPS mass of ungauged supergravity
depending on  the electric charges of the black hole solutions $q_I$ have
to be replaced by the gravitino--gaugino charges (FI terms) $V_I$ to find
the critical points of the BPS mass of the gauged theory. In ungauged
theory one finds the black hole entropy from the BPS mass, in the gauged
theory one finds the cosmological constant. Thus as a function of gravitino
charges $V_I$ and CY
 intersection numbers $C_{IJK}$, the cosmological constant at the
supersymmetric
  critical point
is given by the same topological formula, which defines the entropy of the
extreme supersymmetric black holes:  the value of the cosmological constant
is moduli-independent and depends only on $V_I$ and $C_{IJK}$.

 Thus all previous studies of CY black
hole entropy may be used for understanding the $adS_5$ vacua. In what
follows we will focus our attention on  the issue of non-uniqueness of
supersymmetric critical points of this theory.

Simultaneously with the study of the critical points of the central charge,
one has to verify that at the given critical point some natural physical
conditions are satisfied. We will try to find  fixed points where both the
metric of the moduli space $g_{ij}=-3 C_{IJK} t^I t^J_{,i} t^K_{,k}  $ and
the metric of the vector space (the gauge coupling matrix) $\tilde G_{IJ}=-
   \partial_I \partial_J (\ln {\cal V} ) |_{{\cal V} =1}$
are  positive.\footnote{Negative sign  of kinetic terms in the Lagrangian
usually leads to vacuum instability. However, if one simultaneously changes
the sign  of kinetic and potential energy  of some  fields and if these
fields live only in a part of the  universe different from ours (or on a
different brane) then  one can avoid instabilities. This possibility
deserves investigation because it may provide an explanation of the
vanishing  of the cosmological constant \cite{Linde88}.}

In some cases these conditions are sufficient to guarantee the uniqueness
of the critical point for the black hole entropy. Such cases  give examples
of supersymmetric
 attractors with one basin
of attraction. In some other cases specified by different values of the
intersection numbers $C_{IJK}$ and charges $q_I$ or $V_I$ more than one
critical point satisfying physical conditions may be available.
Particularly in the case of many moduli when the stabilization equations
are higher order algebraic equations, one may expect to find several
critical points consistent with physical requirements. In what follows we
will present conditions for multi basin attractors, give examples  and
discuss the conceptual issues associated with such systems.

We start with an example for $I=1,2$ when the algebraic system common for
d=4 and d=5 theories has a general solution describing d=4 and d=5 black
holes and $adS_5$ vacua. In this example the issue of the uniqueness of the
d=4 black hole entropy and the possibility of the non-uniqueness of the
entropy of d=5 black holes and  the critical points of the $adS_5$ vacua
will be clarified. One may hope to learn some lessons from this simple
system which may help to understand the theories with many moduli.

\

2. Consider the simple case of  $I=1,2$ and generic $C_{IJK}$ and $V_I$. We
choose
\begin{equation}
C_{111}=a\ ,\quad C_{112}=b\ ,\quad C_{122}= c \ ,\quad C_{222}= d \ ,
\end{equation}
and define $\bar t^1\equiv x$ and $\bar t^2\equiv y$. The stabilization
equations consist of a system of two quadratic equations for two variables:
\begin{eqnarray}
ax^2+2bxy+cy^2 =V _1\ ,\label{ek1}\\ bx^2+2cxy+dy^2 =V _2\ .\label{ek2}
\end{eqnarray}
We introduce the following notations\footnote{We assume that $M\neq 0$ and
$L^2 \neq 4MN$.}
\begin{equation}
M \equiv c^2-bd\ , \qquad N \equiv b^2-ac \ , \qquad  L \equiv ad-bc \ ,
\end{equation}
\begin{equation}
{\cal D} \equiv (MV _1^2+ N V _2^2+ L V _1 V _2) \ .
\end{equation}

In the context of black holes and cosmological constant we also introduce
\begin{equation}
E \equiv cq_1 -bq_2\ ,    \qquad F \equiv d q_1 -c q_2 \ ,
\end{equation}
and
\begin{equation}
E \equiv c V_1 -bV_2  \ ,  \qquad F \equiv d V _1 -cV_2 \ ,
\end{equation}
 respectively.

The metric of the moduli space is required to be positive everywhere in the
 moduli space.
  This leads to a requirement that
\begin{equation}
L^2- 4MN<0 \ , \qquad M>0\ , \qquad N>0\ . \label{metric}\end{equation} It
follows from the fact that the expression
\begin{equation}
M \phi^2 - L \phi + N>0
\end{equation}
has to be positive for all values of real $\phi= {y\over x}$.
 This is a natural
condition for the physical theory.  In addition it provides the condition
that the critical points are local minima of the black hole potential and
local maxima of the gauge theory potential \cite{GFK}. We exclude $xy$ and
$y^2$ in favour of $x^2$:
\begin{eqnarray}
2xy= -{F \over M}  +{L x^2\over M}   \ ,\label{xy}\\ \nonumber\\ y^2=
{E\over M}  + {Nx^2\over M} \ . \label{y^2}
\end{eqnarray}
 The system of
equations can now be reduced to the quadratic equation for the variable
$x^2$. The solution of this quadratic equation is:
\begin{eqnarray}
x^2_{\pm} = {FL+2EM \over L^2- 4MN} \pm  { \sqrt{4 M^2 {\cal D} }\over L^2-
4MN} \ ,
\end{eqnarray}
and  $(xy)_{\pm}$ and  $y^2_{\pm} $ are given by (\ref{xy}) and (\ref{y^2})
 in terms of $x^2_{\pm}$, respectively. Thus we have two
sets of solutions of stabilization equations. Let us call them a\, +
critical point and  a\, --  critical point for the $x^2_+,\; y^2_+,\; xy_+$
solutions and
 $x^2_-, \; y^2_-, \; xy_-$ solutions, respectively. The condition
for
 the existence of these solutions is that
 \begin{equation}
{\cal D} \equiv MV _1^2+ N V _2^2+ L V _1 V _2>0 \ .
\end{equation}
One can verify that this condition is satisfied if  the moduli space metric
is positive, i.e. Eq.  (\ref{metric}) is satisfied.
 Notice that $L^2- 4MN<0$ and $ M>0 $, and therefore the second term
 in $x^2_{\pm}$ is always negative:
 \begin{eqnarray}
x^2_{+}  - x^2_{-}  =2  { \sqrt{4 M^2 {\cal D} }\over L^2- 4MN}<0 \ .
\end{eqnarray}
The same situation takes place for the critical values of $y^2_{\pm}$:
\begin{equation}
y^2_{\pm}=\frac{E}{M}+\frac{Nx^2_{\pm}}{M}
=-\frac{HL+2EN}{L^2-4MN}\pm\frac{\sqrt{4N^2{\cal D}}}{L^2-4MN} \ ,
\end{equation}
where $H=bq_1-aq_2$. Here we also find that
 \begin{eqnarray}
y^2_{+}  - y^2_{-}  =2  { \sqrt{4 N^2 {\cal D} }\over L^2- 4MN}<0 \ .
\label{y} \end{eqnarray} This means that we may look for the situation when
$x^2_+$  and $ y^2_+$ are negative and $x^2_-$ and $ y^2_-$ are positive.
This would mean that $x_+$ and $y_+$ are imaginary and $x_-$ and $ y_-$ are
real. This is consistent with the fact that our original variables $t^I =
{\bar t^I\over \sqrt Z}$ are real if
 \begin{equation}
 Z_+ < 0 , \qquad Z_- >0 \ .
\label{sign}\end{equation} The critical values  of the cube of the central
charge are given
 by the following expression (in the black hole case)
\begin{eqnarray}
Z_{cr}(q)^3_{\pm}& =&( t^I_{\pm} q_I)^3= (\bar t^I_{\pm} q_I)^2 =
(xq_1+yq_2)^2_{\pm}\nonumber\\ &= & x^2_{\pm}q_1^2 +y^2_{\pm} q_2^2 +
2xy_{\pm} q_1 q_2 \ .
\end{eqnarray}
Here again we see that it is consistent to have  $Z_+ < 0$ and $Z_-
>0$
for the two critical points. For the cosmological constant case, there is
an analogous expression for the central charge in terms of $V_I$ instead of
$q_I$.

The critical values of the black hole entropy and the critical values of
the  cosmological constant are given by analogous formulae.
 By
substitution of the critical values of $x^2_{\pm}, \; y^2_{\pm}, \;
xy_{\pm}$ we get two critical values of the central charge, entropy and
cosmological constant:
\begin{eqnarray}
Z_{cr}(q)^3_{\pm}  = -{q _2 (d q _1^2+bq_2^2-2cq_1q _2) \over M}+ {\cal D}
\left[{FL +2E M  \pm  \sqrt{4 M^2 {\cal D} }\over M( L^2- 4MN)} \right ],
 \end{eqnarray}

\begin{equation}
\tilde S_{\pm}  =| Z_{cr}(q_I)_{\pm }|^3\ , \qquad (\Lambda_{adS_5})_{\pm}
=- g^2 | Z_{cr}(V_I)_{\pm}|^2 \ .
\end{equation}

Now we have to study the values of  vector space metric (gauge couplings)
at the critical points:
\begin{equation}
\hat  {G}_{IJ}=-
   {Z^2\over 9}  \partial_I \partial_J (\ln {\cal V} ) |_{{\cal V}
=1}= V_I V_J - {2\over 3} C_{IJK} \bar t^K   \bar t^L V_L  \ .
\label{gauge}\end{equation} A tedious calculation, analogous to that
performed
 in  \cite{Marina1} with respect to $d=4$ black holes and moduli
metric,
 allows us to find the generic expression for the
  determinant of the
gauge coupling matrix at the critical point:
\begin{eqnarray}
 (\det\hat G_{IJ})_{\pm}= \mp \frac{2}{9}Z^3_{\pm}  \sqrt{{\cal D}}>0
 \qquad
\rm for \;
 M>0 \ .
\label{det}\end{eqnarray}
 For black holes $Z=q_1x+q_2y$ and for the
cosmological constant it equals $Z=V_1x + V_2y$.

Now we see that one may be able to justify the existence of both critical
points as possibly acceptable  under the following conditions: We have to
require that {\it at the + critical point the critical value of the central
charge $Z_+$ is negative  and at the -- critical point the critical value
of the central charge $Z_-$ is positive}.
 In such case the determinant of  gauge couplings is positive at both
critical points. But this is precisely the condition obtained from the
consistent solution of the attractor equations as shown in (\ref{sign})! In
such cases the determinant of the gauge coupling matrix is positive at both
critical points.

 Note that we  still have to find out whether the
eigenvalues of $\hat G_{IJ}$ are actually positive. It may happen that the
determinant is positive but the eigenvalues are both negative. Fortunately
a completely general treatment is possible in the simple 1-moduli case.
First we note that $\det \hat G_{IJ}= \hat G_{11} \hat G_{22} - \hat
G_{12}^2$ and therefore for all cases where we have established that $\det
\hat G_{IJ}
>0$ we find that
\begin{equation}
\hat G_{11} \hat G_{22}  > \hat G_{12}^2 >0 \ .
\end{equation}
Thus for all cases at hand if we can verify that e.g. $\hat G_{11}$ is
positive, this will guarantee that also $\hat G_{22}$ is positive and that
the total gauge coupling matrix is positive-definite. To derive the
critical values of $G_{11}$ we start with (\ref{gauge}) and get
\begin{eqnarray}
\hat G_{11}= \frac{1}{3}\left(q_1^2\mp 2y^2_{\pm} \sqrt{\cal D}\right).
\label{G11} \end{eqnarray} Now we can  provide a generic answer to the
problem of the positivity of the vector metric. Both critical points are
locally stable  iff $Z_+<0$ and $Z_->0$, as  was already required by the
positivity of the $\det$ of the vector metric. The same condition provides
the positivity of $\hat G_{11}$ since the first term in (\ref{G11}) is
always positive and the second one
 is also positive, as  follows from eqs. (\ref{y}), (\ref{sign}).
Thus
  {\it the moduli space metric and the gauge coupling matrix are
both positive
   definite iff}
\begin{equation}
L^2- 4MN<0 \ , \qquad M>0\ , \qquad N>0\ ,   \qquad Z_+ < 0\ , \qquad Z_-
>0 \ , \label{conditions}\end{equation} and one can claim that there are
two {\it critical points that form the stable local minima of the entropy
of d=5 black holes and of the absolute value of the cosmological constant
of the $adS_5$ vacua}.

These are generic conditions in the one-moduli  problem under which two
physically acceptable critical points are possible. One can give many
numerical examples of cases where  these conditions are satisfied. Consider
for example the case suggested in  \cite{BC} which in our notation is
$a=0$, $b=2$, $c= 2.8  \sqrt 3$, $d=6$, $V_1=1$, $V_2 = -  0.6 \sqrt 3$. It
has been verified  in \cite{BC} that the moduli space metric is positive.
However the vector space metric was not studied. As it is now clear from
our studies, the vector space metric for this example is indeed positive
since the central charge has an opposite sign in two critical points.

An interesting comment can be made about the properties of the excitations
around the two gauged theory vacua. The gravitino mass near one critical
point is positive $M_{-grav} = Z_- >0$ and the one near the second critical
point is negative $M_{+ grav} = Z_+<0$ since its value at each critical
point is the value of the central charge. The issue of the positive versus
negative fermion mass in $d=5$ was discussed in \cite{witten}, where it was
noticed that both the positive mass theory and the negative mass theory
may exist. The relations between these theories include a change of
$\gamma^\mu $ matrices into $-\gamma^\mu$, as well as of the representation
of the little part $SO(4)$ of the Lorentz group. Interestingly, we need
both versions of the theory to make acceptable not only the vacuum state
but also the excitations around each vacuum.

The conditions for more than one basin of attraction can be easily
violated, in which case the system has only one attractor point. For
example we may have some $C_{IJK}$ and $q_I$  (or $V_I$) for which either
the solution is unique or the central charge has the same sign at both
critical points. This would mean that one of the critical points is
unstable with respect to the generation of vector fields and has to be
excluded.

Typically, the values of $Z^2$ at two critical points is different. An
interesting question therefore here   is: Is the solution with $Z_+<0$ and
$Z_->0$ and $|Z_+|= |Z_-|$ possible? The equation for our parameters
specifying this case is:
\begin{eqnarray}
{\cal A} \equiv -V _2 (d V _1^2+bV_2^2-2cV_1V _2) +  {\cal D} \left[{FL +2E
M \over ( L^2- 4MN)} \right]=0 \ .
 \label{cosm} \end{eqnarray}
We have found 3 families of solutions for such configurations. We take the
following set of parameters $a=0,~ b=1/3, ~ d=1, ~V_1=1$ and we do not
specify $c$ and $ V_2$, except that $c^2> {4\over 9}$. Equation ${\cal
A}=0$ has
 three types of  solutions: $V_2= {3\over 2}c$ or  $V_2=  {3\over 2}[
c \pm \sqrt{9c^2 -4}]$.  The physical conditions of the positivity of the
moduli space metric and of the gauge coupling matrix are satisfied for any
of these solutions, for all $c^2>4/9$. For example, $a=0,~ b=1/3,~ c=4/3,~
d=1, ~V_1=1,~ V_2=2$ gives a consistent set.

\

3.  Having established the existence of two supersymmetric critical
 points describing the $adS_5$ vacua  some of which may have equal cosmological
  constant we would like to find also the interpolating domain wall solution.
Here we follow \cite{ST,gary} and take an ansatz
\begin{equation}
ds^2= e^{2A(r)} dx^\mu dx^\nu \eta_{\mu\nu} + dr^2 = U^2 dx^\mu dx^\nu
\eta_{\mu\nu}
 + {1\over (\partial_r A)^2} \,  {dU^2\over U^2} \ ,
\label{metric1}
\end{equation}
where $U=e^A$. At the critical points where $(\partial_r
A)^2_{cr}=Z^2_{cr}
 $ the geometry is an $adS_5$ space with a cosmological constant
 $\Lambda = -6 Z^2_{cr}$.
The equations of motion of the gauged supergravity describing a domain wall
can be derived from the energy functional:
\begin{equation}
E= {1\over 2} \int_{\infty}^{+\infty} dr e^{4A} \left \{ [ g^{1/2} \phi'
\mp 3  Z_{, \phi} g^{-1/2}]^2 - 12 [  A' \pm Z]^2 \pm 3 [ e^4 A Z]
^{+\infty}_{-\infty} \right \}  .
\end{equation}
Here $\phi'\equiv  \partial _r \phi, A'\equiv  \partial _r A$. An analogous
expression was also presented  in \cite{ST,gubser} where it was also
noticed that as different from the standard BPS situation one of the
squares
 enters with the negative sign. Our energy functional has a non-trivial
  moduli space metric
  $g= g_{\phi \phi}$ which is absent in \cite{ST,gary,gubser}.
This term is important because it provides a possibility to obtain a smooth
supersymmetric solution for $\phi(r)$ interpolating between the two
different vacua. The first order equations of motion
   of the gauged supergravity which admit Killing spinors are
   \cite{BC,ST,gary,gubser}
\begin{equation}\label{qq}
\phi' (r) =  \pm 3 g^{-1} Z_{, \phi}\ , \qquad \qquad  A' = \mp Z \ .
\end{equation}
We  solved these equations for a wide variety of parameters which allow
existence of two attractors with equal values of the cosmological constant
but opposite values of $Z$, in a hope to find a domain wall solution of
Randall-Sundrum type, with $A  \propto -|r|$ at large $|r|$. However,
instead of that we always found solutions\footnote{According to eq.
(\ref{metric1}), the solutions with $A \sim {|r|}$ tend to the large $U$,
 i.e. to UV, whereas  those with $A \sim {-|r|}$  tend to small $U$, i.e.
 to IR.} with $A \propto  |r|$, see one of these solutions presented in Fig.
1.

\FIGURE{\epsfig{file=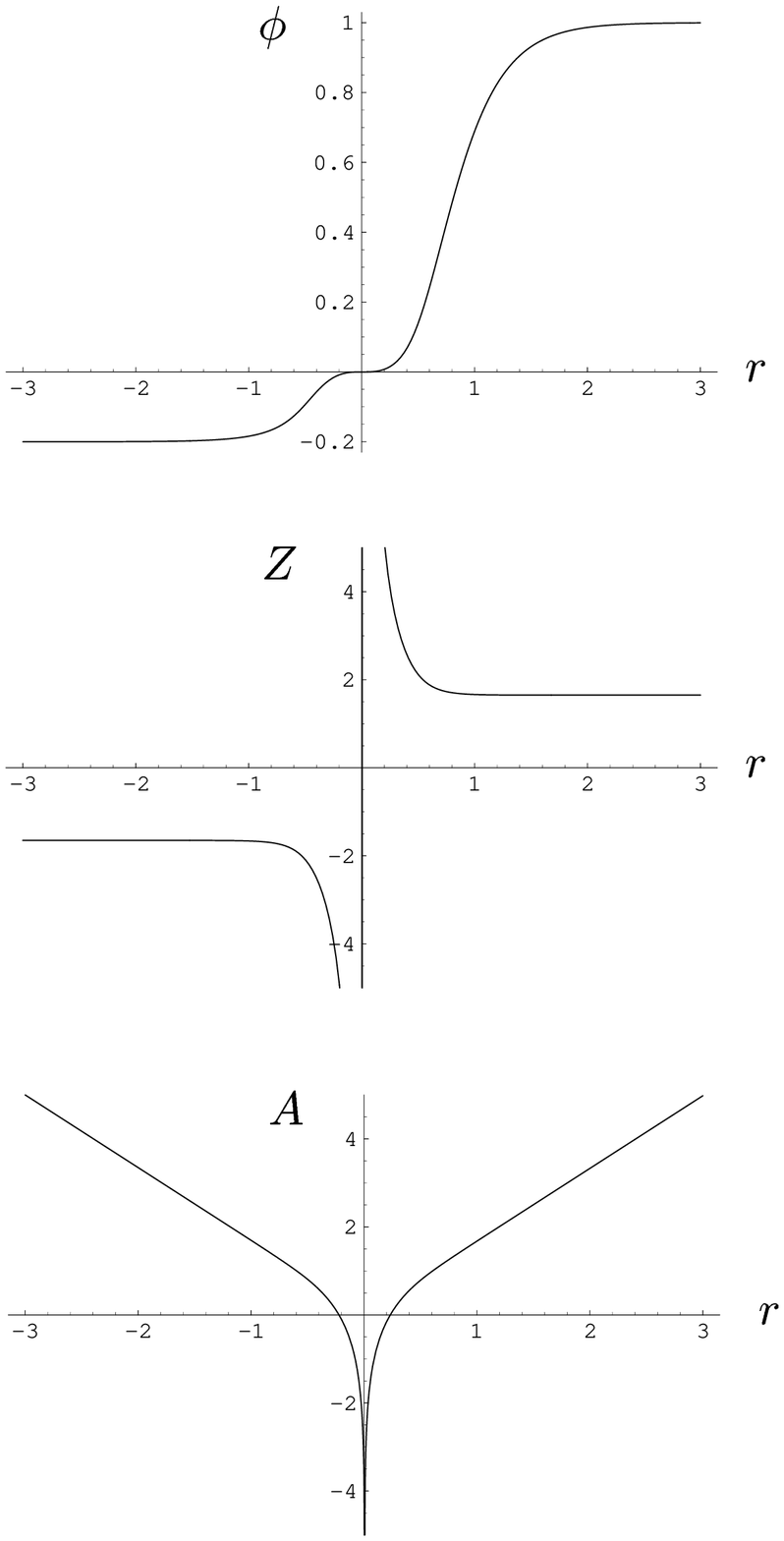,width=10cm}
        \caption{A solution for the scalar field $\phi$
        interpolating between
        two different vacua with equal values of $|Z|$.
        Note that $\phi(r)$ is nonsingular because
        of the vanishing of $g^{-1}$ at $\phi = 0$, whereas $Z$
        and $A$ are singular: $Z \sim r^{-1}$ and $A(r) \sim \log |r|$ at $|r| \to 0$. At
large $r$ the function
        $A(r)$ grows as $|r|$ rather than decreases as $-|r|$.
         This is a general property of interpolating
         solutions
          in our class of models.}
    \label{myfigure}}

One can show that this is a general result, which follows from the fact
that, according to \cite{CKRRSW}, one has $\partial_i \partial_j Z =
{2\over 3} g_{ij} Z$ at the supersymmetric critical point. Consider, for
example, the case $ \phi' (r) =  - 3 g^{-1}    Z_{, \phi}$ in Eq.
(\ref{qq}). The scalar field  trajectory can approach the critical value at
large $r$ either from below or
    from above. The one from above with the negative derivative of $\phi$   requires
    that
     $Z_{, \phi} >0$ above the attractor. The one from below requires that
      $Z_{, \phi} <0$
     below the attractor, whereas the attractor point corresponds to $Z_{,
     \phi}=0$, therefore
      the second
      derivative of the central charge is positive,  $Z_{, \phi \phi }>0$. Since we
       consider the
       situations where $g_{ij}>$ in the attractor, equation $\partial_i \partial_j Z =
       {2\over 3} g_{ij} Z$ implies that  the right
      critical point has a positive central charge $Z$. Then
      equation $ A' = + Z$ implies
      that $A'$ approaches a constant positive value at large positive
      $r$, i.e.   $A \sim |Z_{cr}r|+ const$, see Fig. 1. If we would have used the second pair
of equations, staring with $ \phi' (r) = 3 g^{-1}    Z_{, \phi}$  we would
find that
 at the right critical point $Z$ is
negative, but in this case $ A' = - Z $ and again  $ A'
>0$.  The case when $\partial_i \partial_j Z =
{2\over 3} g_{ij} Z=0$ would lead to $A'=0$ and is also ruled out.
 Thus it follows from  supersymmetry that the
interpolating solution which admits
 Killing spinors at large positive $r$  behaves as $A\sim |Z_{cr}r|+const$ at large
 $|r|$. This is not an interpolating domain wall of \cite{RS} as was already
  observed in \cite{BC}.

In \cite{gubser} a non-supersymmetric choice of  $Z$ was suggested in the
framework of supergravity equations. In this way a smooth solution modeling
branes with
 the desired asymptotic
 of the interpolating solution was obtained. One can verify that the choice of the
  function $Z$  in \cite{gubser} is such that $ Z_{,\phi \phi} \sim - Z$ at
  the critical points
  $Z_{,\phi}=0$.
  This condition cannot be valid in a supersymmetric theory where
$\partial_i \partial_j Z =
       {2\over 3} g_{ij} Z$ at the critical points and the moduli
 space metric is positive. This example
 confirms
 that solutions  $A \propto -|r|$ require  violation of supersymmetry in this
       class
       of theories.

Until now  we were looking only for supersymmetric interpolating solutions,
and found that they do not behave as $A \propto {-|r|}$. One may wonder
whether one
 can find more general, non-supersymmetric interpolating solutions with the desirable
  asymptotic $A \propto {-|r|}$.  The answer to this question is also negative. Indeed,
  the relevant
   equation of motion for the interpolating scalars in the background
   metric is
\begin{equation}\label{deviation}
 \phi'' + \left (4 A'+{g_{,\phi}\over g} \phi'\right) \phi' + 6 g^{-1}  P_{,\phi}=0\ ,
\end{equation}
where at the critical points $ P_{,\phi \phi}$ is positive. Let us
assume that the solution of this equation asymptotically approaches an
attractor point $\phi_{cr}$ at large $r > 0$, so that $g$ and $g_{,
\phi}$ become constant, $A'$ becomes  negative constant, and $\phi'$
gradually vanishes at large $r$. Then the deviation $\delta \phi$ of
the field $\phi$
  from its asymptotic value $\phi_{cr}$ at large $|r|$ satisfies the
  following equation:
\begin{equation}\label{deviation2}
 \delta\phi'' - 4 |A'| \delta\phi' = - 6 | g^{-1} P_{,\phi \phi}|\delta\phi  \ .
\end{equation}
This is equation for a harmonic oscillator with a negative friction
 term $-  |A'| \delta\phi'$. Solutions of this equation describe oscillations
 of $\delta\phi$ with
 amplitude blowing up at large $|r|$, which contradicts our
 assumptions.
This argument shows that there are no interpolating solutions in our theory
with $A \propto {-|r|}$ at large $|r|$. This conclusion remains valid even
if one relaxes our assumption that the scalar field metric is positive and
considers domains with $g<0$.

Finally, we would like to note that even though we called the field
configuration shown in Fig. 1 ``interpolating solution,'' its physical
interpretation requires further investigation. Indeed, even though the
solution for the scalar field $\phi$ smoothly interpolates between the
two attractor solutions, the function $A(r)$ is singular. It behaves as
$\log |r|$ at $|r| \to 0$. Metric near the domain wall is given by
\begin{equation}
ds^2= r^2 dx^\mu dx^\nu \eta_{\mu\nu} + dr^2 \ . \label{metric2}
\end{equation}
This implies the existence of the curvature singularity at $r=0$, which
separates the universe into two parts corresponding to the two
different attractors.

\

4. We would like to point out that the existence of several basins of
attraction in dynamical systems in general is quite common. Typically the
system is attracted to the nearest attractor point after it reaches a given
basin of attraction. The new result established in this paper is that there
are conditions when more than one critical points in supersymmetric
attractors are physically acceptable, i.e. the moduli space metric and the
gauge couplings are positive. The system may be at some initial value of a
moduli either from one side of the discontinuity of the moduli space metric
or on the other side. This gives a precise definition of the basin of
attraction. Note that the value of the entropy (or the value of the
cosmological constant) in general is different for two critical points
under discussion:
\begin{eqnarray}
\delta \tilde S =  \left |{\cal A} +2{{\cal D}^{3/2}\over ( L^2-
4MN)}\right |^{3/2}- \left |{\cal A} -2 {{\cal D}^{3/2} \over ( L^2-
4MN)}\right |^{3/2} \ .\label{entropy}\end{eqnarray}
 Thus, with respect
to black holes, our analysis seems to develop and confirm the idea
suggested by Moore \cite{moore} that the black holes (in $d=5$, under some
conditions specified in this paper) may represent a  multiple attractor
system. To specify a black hole one has to specify not only the charges and
the prepotential but also the attractor point, defined by the values of
moduli at some distance from the horizon. However, to fully understand the
issue of the possible non-uniqueness of the black hole entropy one has to
study the
 black hole solutions and not only the critical points.  We hope
 to investigate this in the future.

With respect to $adS_5$, we
 have shown that the critical points for the cosmological constant
correspond to a multiple attractor system. One of the most interesting
issues is related to the ${\bf Z}_2$-symmetric BPS critical points with
equal values of the $adS_5$ radius found in this paper when ${\cal
A}=0$. Our investigation of the domain wall solution in gauged
supergravity \cite{GST} with one vector multiplet shows that for the
non-compact 5-th dimension the asymptotic form of the interpolating
solutions is always $e^{|Z_{cr}r|}$, which has an opposite sign
compared to the Randall-Sundrum scenario \cite{RS,BC}.

The main result of our paper is that multiple basins of attraction are
possible in supersymmetric theories. We found  double-attractor systems
with positive scalar and vector metric  in $d=5$ one-moduli theory. We
expect that in theories with many moduli in $d=4$ as well as in $d=5$ one
may also find
 physically acceptable   configurations with  multiple basins of attraction.

\vskip 1 cm

The authors are grateful to G. Gibbons, M. Gunaydin, S. Ferrara, G. Moore
and J. Rahmfeld for valuable discussions. This work was supported in part
by NSF grant PHY-9870115.


\begin{thebibliography}{30}

\bibitem{FKS}
S.~Ferrara, R.~Kallosh, and A.~Strominger, {\it N=2 Extremal Black Holes},
Phys.\ Rev.\ {\bf D52}, 5412 (1995), hep-th/9508072.


\bibitem{S} A. Strominger, {\it Macroscopic Entropy of $N=2$
Extremal Black Holes},
  Phys.\ Lett.\ {\bf B383}, 39 (1996),
hep-th/9602111.

\bibitem{FK}
S.~Ferrara and R.~Kallosh, {\it Supersymmetry and Attractors}, Phys.\ Rev.\
{\bf  D54}, 1514 (1996), hep-th/9602136;\\ S.~Ferrara and R.~Kallosh, {\it
Universality of Supersymmetric Attractors},
 Phys. Rev.  {\bf D54}, 1525 (1996), hep-th/9603090.

\bibitem{GFK}
 S.~Ferrara, G.~W. Gibbons,  and R.~Kallosh, {\it
 Black Holes and Critical Points in Moduli Space},
Nucl.\ Phys.\ {\bf B500}, 75 (1997), hep-th/9702103.


\bibitem{GST} M.Gunaydin, G. Sierra, and P.K. Townsend,
 {\it Gauging The D = 5 Maxwell-Einstein Supergravity
  Theories: More on Jordan Algebras},
Nucl.\ Phys.\ {\bf B253}, 573 (1985).


\bibitem{CKRRSW} A. Chou, R. Kallosh, J. Rahmfeld, S.-J. Rey, M.
Shmakova, and W.K. Wong, {\it Critical Points and Phase Transitions in 5d
Compactifications of M-Theory}, Nucl.\ Phys.\ {\bf B508}, 147 (1997),
hep-th/9704142.



\bibitem{moore} G. Moore, {\it Arithmetic and Attractors},
hep-th/9807087.


\bibitem{Marina} M. Shmakova,
{\it Calabi-Yau Black Holes}, Phys.\ Rev.\ {\bf D56}, 540 (1997),
hep-th/9612076.

\bibitem{Marina1}  M. Shmakova, PhD thesis, Ch. 4.5,
http://www.slac.stanford.edu/$\sim$shmakova.

\bibitem{BC} K. Behrndt and M. Cveti\v c,
{\it Supersymmetric Domain-Wall World from D=5 Simple Gauged Supergravity},
hep-th/9909058.


\bibitem{malda} J. Maldacena,
{\it The Large N Limit of Superconformal Field Theories and Supergravity},
Adv.\ Theor.\ Math.\ Phys.\ {\bf 2}, 231 (1998), hep-th/9711200.

\bibitem{kelly}
A.~Lukas, B.A.~Ovrut, K.S.~Stelle and D.~Waldram, {\it The Universe as a
Domain Wall}, Phys.\ Rev.\ {\bf D59}, 086001 (1999) hep-th/9803235.


\bibitem{RS} L. Randall and R. Sundrum,
{\it A Large Mass Hierarchy from a Small Extra Dimension},
hep-ph/9905221;\\
  L. Randall and R. Sundrum,
{\it An Alternative to Compactification}, hep-th/9906064.


\bibitem{B} K. Behrndt, {\it Domain Walls of D = 5 Supergravity
and Fixpoints of N = 1 Super Yang- Mills}, hep-th 9907070.



\bibitem{ST} K. Skenderis and P.K. Townsend,
{\it Gravitational Stability and Renormalization-Group Flow}, hep-th
9909070.


\bibitem{gary}
A.~Chamblin and G.W.~Gibbons, {\it Supergravity on the Brane},
hep-th/9909130.


\bibitem{gubser}
O.~DeWolfe, D.Z.~Freedman, S.S.~Gubser and A.~Karch, {\it Modeling the
Fifth Dimension with Scalars and Gravity}, hep-th/9909134.

\bibitem{hawking}
A.~Chamblin, S.W.~Hawking, and H.S.~Reall, {\it Brane-World Black Holes},
hep-th/9909205.

\bibitem{Linde88} A.D. Linde, {\it The Universe Multiplication and
the Cosmological Constant Problem}, Phys.\ Lett.\ {\bf B200}, 272 (1988).




\bibitem{witten} E. Witten, {\it Phase Transitions in M-Theory and
F-Theory}, Nucl.\ Phys.\ {\bf B471}, 195 (1996), hep-th/9603150.




\end{thebibliography}
\end{document}